\documentclass{PoS}

\usepackage{url}
\usepackage{subfig}

\newcommand{\myUrl}[1]{\href{{#1}}{{#1}}}

\def\cell{PowerXCell~8i}
\def\celcius{\mbox{}{\rm ^oC}}

\title{QPACE -- a QCD parallel computer based on Cell processors}

\ShortTitle{QPACE}

\author{%\Large
  H.~Baier$^1$,
  H.~Boettiger$^1$,
  M.~Drochner$^2$,
  N.~Eicker$^{2,3}$,
  U.~Fischer$^1$,
  Z.~Fodor$^3$,
  A.~Frommer$^3$,
  C.~Gomez$^4$,
  G.~Goldrian$^1$,
  S.~Heybrock$^5$,
  D.~Hierl$^5$,
  M.~H\"usken$^3$,
  T.~Huth$^1$,
  B.~Krill$^1$,
  J.~Lauritsen$^1$,
  T.~Lippert$^{2,3}$,
  \speaker{T.~Maurer}$^5$,
  B.~Mendl$^5$,
  N.~Meyer$^5$,
  A.~Nobile$^5$,
  I.~Ouda$^6$,
  M.~Pivanti$^7$,
  \speaker{D.~Pleiter}$^8$,
  M.~Ries$^1$,
  A.~Sch\"afer$^5$,
  H.~Schick$^1$,
  F.~Schifano$^7$,
  H.~Simma$^{8,9}$,
  S.~Solbrig$^5$,
  T.~Streuer$^5$,
  K.-H. Sulanke$^8$,
  R.~Tripiccione$^7$,
  J.-S.~Vogt$^1$,
  T.~Wettig$^5$,
  F.~Winter$^8$ \\
  \llap{$^1$}	IBM Deutschland Research \& Development GmbH,
		71032 B\"{o}blingen, Germany\\
  \llap{$^2$}	Research Center J\"{u}lich, 52425 J\"{u}lich, Germany\\
  \llap{$^3$}	University of Wuppertal, 42119 Wuppertal, Germany\\
  \llap{$^4$}	IBM La Gaude, Le Plan du Bois, La Gaude 06610, France\\
  \llap{$^5$}	Department of Physics, University of Regensburg,
		93040 Regensburg, Germany\\
  \llap{$^6$}	IBM Rochester, 3605 HWY 52 N, Rochester MN 55901-1407, USA\\
  \llap{$^7$}	University of Ferrara, 44100 Ferrara, Italy\\
  \llap{$^8$}	Deutsches Elektronen Synchrotron (DESY),
		15738 Zeuthen, Germany\\
  \llap{$^9$}	Department of Physics, University of Milano-Bicocca,
		20126 Milano, Italy
}

\abstract{%
QPACE is a novel parallel computer which has been developed to be
primarily used for lattice QCD simulations. The compute power is
provided by the IBM {\cell} processor, an enhanced version of the
Cell processor that is used in the Playstation 3. The QPACE nodes are
interconnected by a custom, application optimized 3-dimensional torus
network implemented on an FPGA. To achieve the very high packaging density
of 26 TFlops per rack a new water cooling concept has been developed
and successfully realized. In this paper we give an overview of the
architecture and highlight some important technical details of the system.
Furthermore, we provide initial performance results and report on the
installation of 8 QPACE racks providing an aggregate peak performance of
200~TFlops.
}

\FullConference{%
  The XXVII International Symposium on Lattice Field Theory - LAT2009\\
  July 26-31 2009\\
  Peking University, Beijing, China
}

\begin{document}

%===============================================================================
\section{Introduction}

In the past custom-designed supercomputers have contributed a significant
fraction of the compute cycles available for lattice QCD calculations.
They did not only make compute power available at affordable
costs, they also ensured the availability of highly scalable architectures
needed for many calculations. However, design, production and deployment of
such cost-efficient machines becomes more challenging with prices per
GFlops for commodity systems going down and technology becoming more
complex. For example, the latest generation of custom machines,
apeNEXT~\cite{apenext} and QCDOC~\cite{qcdoc}, was based on custom designed
processors.  Given the costs and the risks involved in an ASIC design, this
has become less of an option today.

% Concept
In the QPACE project we therefore decided to explore a different
concept.  Instead of a custom designed processor we select fast
commodity processors which we interconnect via a custom network.
This network is implemented through a companion processor
which we call the \emph{network processor} (NWP) and for which we use a
re-configurable Field Programmable Gate Array (FPGA) instead of an ASIC.
The nodes consisting of a commodity processor and a network processor then have
to be integrated into a cost-efficient system design.

% Processor requirements and selection
To realize such a concept a suitable commodity processor has to be selected.
Achieving good performance for key lattice QCD kernels is an obvious
criterion. Using a powerful multi-core processor helps to reduce the
required number of nodes (which is likely to reduce costs) and relax the
requirements for the network bandwidth. One also has to take into account
that connecting the NWP to the I/O interface of a complex processor is a
technical challenge.  At the time of starting the project we identified
the IBM {\cell} processor as a suitable candidate.

% Network requirements
For lattice QCD applications to scale on a large number of processors
it is not only the network bandwidth that matters. For typically used
algorithms, data are communicated using relatively small messages, and
the network latency will thus have a major impact on the efficiency
that can be achieved.  Frequently used custom network technologies
(e.g., Infiniband) provide a bandwidth which exceeds our requirements,
which for the given processor we estimated to be $O(1)$~GBytes/sec
per node and direction. Latencies, however, are typically more than
$O(10)$~$\mu$sec. On the {\cell} running at 3.2~GHz this corresponds to
$O(30,000)$ clock cycles. Using custom-designed hardware and software
our goal was to bring this down to $O(1)$~$\mu$sec.

% System requirements + TCO
A final challenge in terms of system design is the cost-efficient
integration of the basic building blocks into a system.  While development
and procurement costs still constitute the largest fraction of the total
cost of ownership, operational costs, in particular costs for electricity
and cooling, are becoming more and more relevant. Indeed, power-efficient
computing has become a major topic in the area of HPC.

% Organisation of the paper
After having outlined the requirements and challenges of the QPACE project
we will in the next section, section~\ref{section:cbe}, discuss the processor in more
detail. In the following sections~\ref{section:architecture}
and~\ref{section:hardware} we will provide an overview of the architecture
and highlight some of the hardware components designed in this project.
In section~\ref{section:nwp} we will focus on one of the core components,
the network processor, and its implementation on an FPGA.
Section~\ref{section:tnw} describes the concept of the torus network
and provides results from some initial performance measurements.
We then continue by highlighting the novel cooling system and the power
efficiency of the system in section~\ref{section:system}.
Section~\ref{section:frontend} explains how the system is integrated with
a front-end system. Section~\ref{section:appl} provides
an overview of initial experiences implementing application programs for QPACE.
Before presenting our conclusions in section~\ref{section:conclusion} we
briefly describe the organization of this project in
section~\ref{section:prjorg}.

%===============================================================================
\section{{\cell} processor}\label{section:cbe}

The {\cell} processor is a more recent implementation of the
Cell Broadband Engine Architecture~\cite{cbea}. A first implementation
has been developed by Sony, Toshiba and IBM with the first major
commercial application being Sony's PlayStation 3. The {\cell}
is an enhanced version of that processor with support for high-performance
double precision operations, IEEE-compliant rounding, and a DDR2 memory interface.

\begin{figure}[ht]
\hspace*{3mm}
\subfloat[{\cell} processor]{
  \label{fig:cbe}
  \includegraphics[scale=0.325]{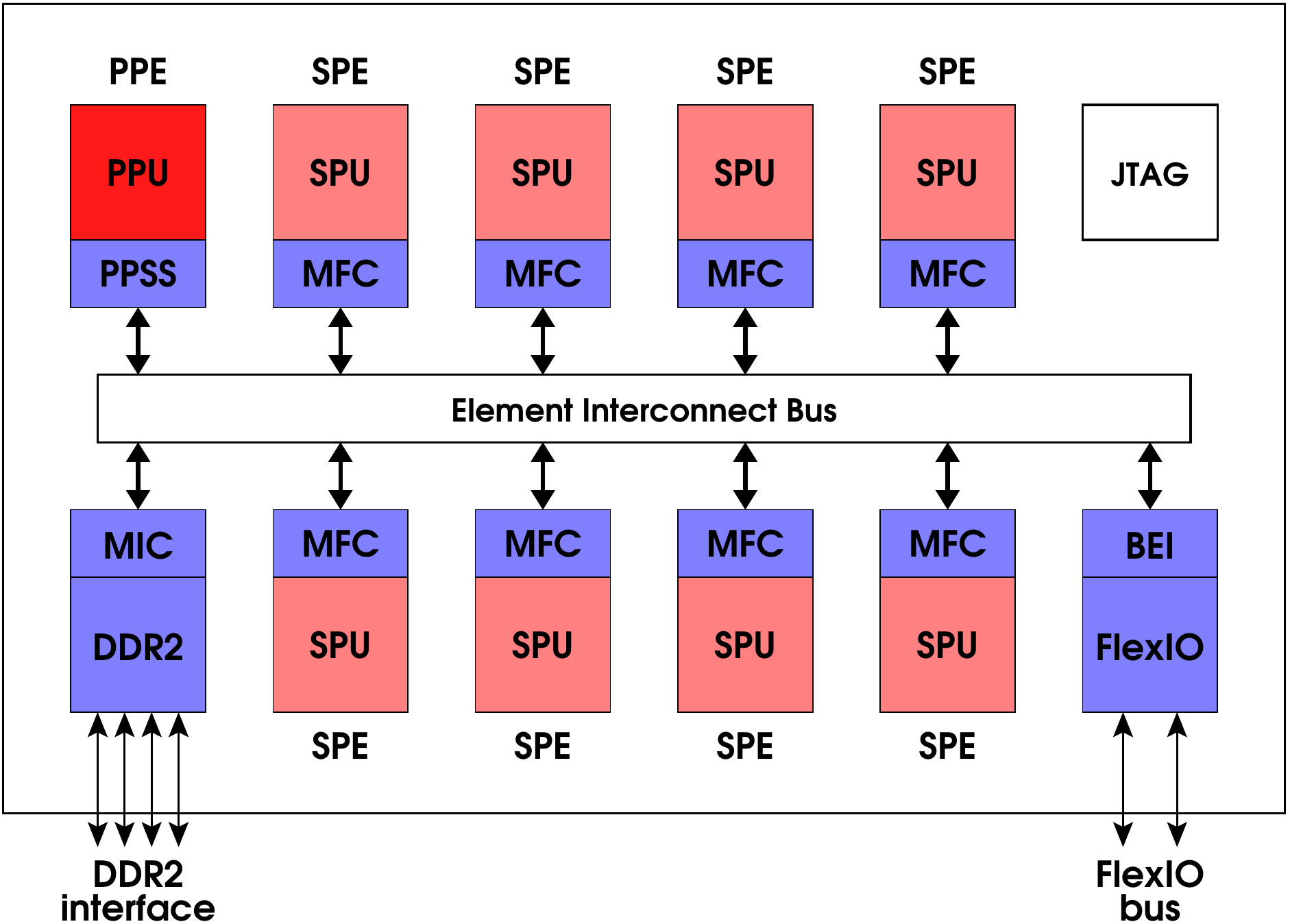}
}
\hfill
\subfloat[QPACE Node card]{
  \label{fig:rialto}
  \raisebox{1mm}{\includegraphics[scale=0.29]{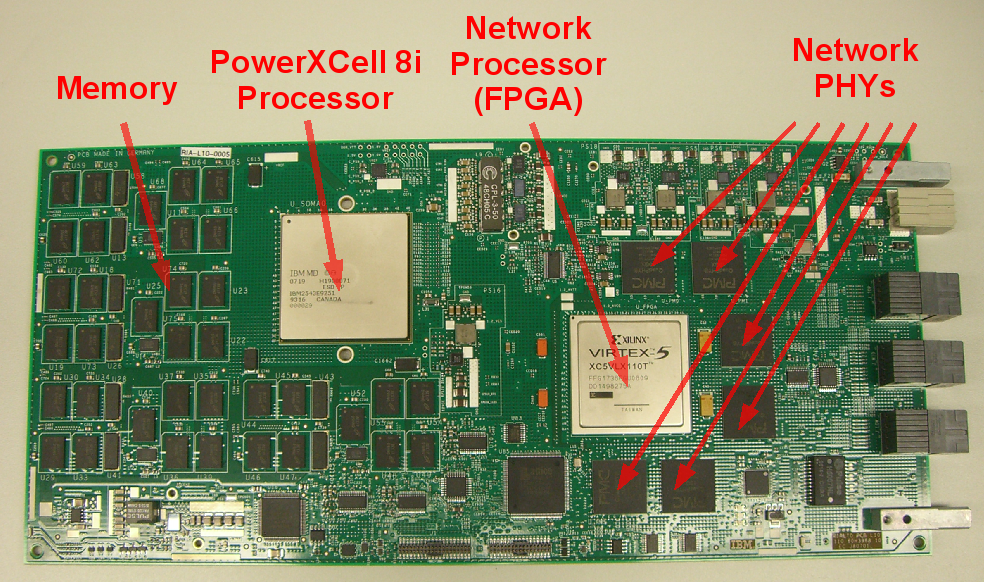}}
}
\hspace*{3mm}
\caption{The left panel shows a schematic overview of the {\cell} processor
with 8 Synergistic Processing Elements (SPE) and a Power Processing Element
(PPE). The other components are described in the text.
The right panel shows the QPACE node card.}
\end{figure}

The {\cell} comprises multiple cores (see Fig.~\ref{fig:cbe}) including
the Power Processing Element (PPE), which is a standard PowerPC core
that can, e.g., be used for running the operating system Linux. From there
threads can be started on the 8 Synergistic Processing Elements (SPE).
These cores as well as the Memory Interface Controller (MIC) and the
Broadband Engine Interface (BEI) are interconnected via the Element
Interconnect Bus (EIB). This bus has a very high bandwidth of up to
200 GBytes/sec. The JTAG interface provides access to processor
registers for debugging.

The real compute power is provided by the 8 SPEs. Each of them consists of
a Synergistic Processing Unit (SPU) and a Memory Flow Controller (MFC). In
each clock cycle the SPU can perform a multiply-add operation on a vector
of 4 (2) single-precision (double-precision) floating-point numbers.
This gives a peak floating-point performance per SPU of 25.6 or
12.8~GFlops in single or double precision, respectively. With a peak
performance of 205 (102) single-precision (double-precision) GFlops
the {\cell} was the most powerful commercially available processor
at the start of the QPACE project.  The same processor is also used in the
Roadrunner supercomputer~\cite{roadrunner}, which at the time of writing
this paper is still number one on the Top500 list~\cite{top500}.

The memory hierarchy of the processor is non-trivial. Each of the SPUs
has its own Local Store (LS), a 256 kBytes on-chip memory from which
up to one 16 Byte word per clock cycle can be loaded or stored to or from the
SPU's register file.  Loading and storing data from and to other devices
attached to the EIB is handled by the Direct Memory Access (DMA) engine
in the MFC. The interface to the external memory, the MIC, provides a
bandwidth of 25.6~GBytes/sec. This interface is shared by all SPEs, thus
reducing the bandwidth per clock cycle and SPE to 1 Byte peak.

From a naive comparison with the bandwidth requirements of the most
important application kernel, the fermion matrix-vector multiplication,
this bandwidth is small compared to the floating-point performance. Let
us consider the case of Wilson-Dirac fermions, where the ratio of
single-precision Flops per Byte which has to be loaded or stored is
$0.9$. Comparing the SPU single-precision performance and the bandwidth
between register file and LS we have a ratio of 8~Flops/16~Bytes, i.e., the
bandwidth exceeds our requirements. But for loads from and stores to main
memory this ratio changes to 64~Flops/8~Bytes.  This simple comparison
of hardware characteristics and one of the important application
performance signatures indicates that it is possible to achieve a good
sustained performance if implementation is heavily optimized with respect
to reducing memory access and keeping data in the LS.  For a more detailed
performance analysis see~\cite{lat07,cise,nobile}.

%===============================================================================
\section{Architecture overview}\label{section:architecture}

\begin{figure}[ht]
\hspace*{10mm}
\begin{minipage}{60mm}
\begin{center}
\subfloat[Rack]{
  \label{fig:rack}
  \includegraphics[scale=0.335]{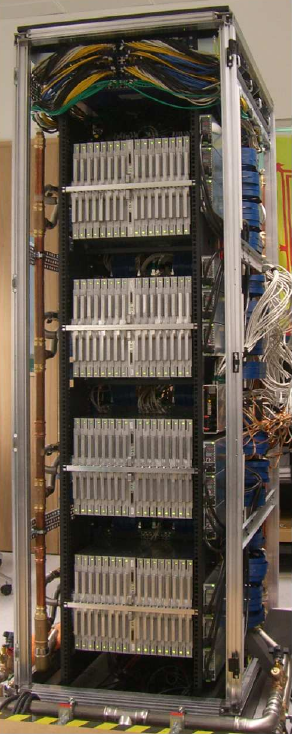}
}
\end{center}
\end{minipage}
\hfill
\begin{minipage}{60mm}
\begin{center}
\subfloat[Superroot card]{
  \label{fig:src}
  \includegraphics[scale=0.250]{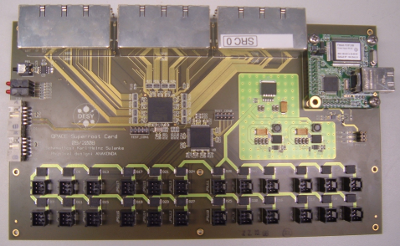}
}\\
\subfloat[Backplane]{
  \label{fig:backplane}
  \includegraphics[scale=0.220]{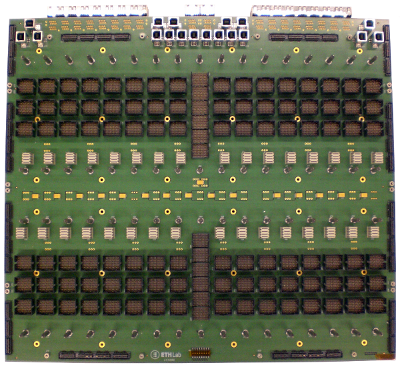}
}
\end{center}
\end{minipage}
\hspace*{10mm}
\caption{The left panel shows a QPACE rack during assembly in the IBM
B\"oblingen lab. Visible are 4 units with 32 node cards each. The vertical
copper pipes at the left side are part of the cooling system.
Pictures (b) and (c) show a QPACE superroot card and a backplane.
} 
\end{figure}

% Component overview
The smallest building block of QPACE is the \emph{node card}. Node cards
are inserted into \emph{backplanes}, each of which can host up to 32
node cards plus 2 \emph{root cards}. The latter can be used to monitor
and control up to 16 node cards mounted in the same row.  One QPACE
\emph{rack} can house up to 8 backplanes, i.e., the maximum number of
node cards per rack is 256. In Fig.~\ref{fig:rack} only 4 units with 32
node cards each are visible. Due to the \emph{water-cooling sub-system}
there is no need for air being able to flow from one side of the rack
to the other.  Therefore the other side of the rack can be used to
mount another 128 node cards.

% Power distribution
Power is distributed via the backplanes to node and root cards.
Three hot-swappable power supply units (PSU) are attached to each
backplane, one of them being redundant.  To connect the backplane and the
PSU we designed a \emph{PSU interface board} which makes it possible
to access the PSUs via a serial I2C link. These links connect all PSUs
within one rack to a \emph{superroot card}
which is the central PSU monitoring and control instance.

% Networks
The QPACE architecture comprises 3 different types of networks:
\begin{itemize}
\item The \emph{torus network} is a low-latency, high-bandwidth
      network. This network connects each node card to its 6 nearest neighbors
      within a torus. We will describe this network in more detail in
      section~\ref{section:tnw}.
\item The \emph{Ethernet network} connects node cards as well as the
      root and superroot cards to the front-end system. It is used for
      booting and I/O. (See section~\ref{section:frontend} for more details.)
\item The \emph{global signal network} is a simple 2-wire tree network.
      It can be used for fast evaluation of global conditions and
      synchronization of the node cards as well as distribution of an
      interrupting kill signal to all node cards of a particular partition.
\end{itemize}

% Global clock tree
A global clock signal is distributed via the root cards. This clock
is used to generate the reference clock for the torus network. The main
purpose of this global clock tree is to maximize alignment of the clocks
at transmit and receive side.

% Switches
Also integrated in the rack are 7 \emph{Ethernet switches} to which all
256 node cards, 16 root cards and the superroot card are connected.

%===============================================================================
\section{Hardware details}\label{section:hardware}

% Nodecard
The QPACE node card is shown in Fig.~\ref{fig:rialto}.
The left part of the node card contains the DDR2 memory chips and the {\cell}.
Each node card provides 4~GBytes of main memory. The number of memory
modules has been optimized for maximum bandwidth. A small microcontroller
is used as a Service Processor (SP). This controller not only monitors and
manages the {\cell}, it also controls the on-board voltages and watches
the temperatures. If voltage or temperature thresholds are exceeded
the node card is automatically switched off.

The components sitting on the right half of the PCB are special for the QPACE
architecture, i.e., the FPGA and a set of devices, called PHYs, which implement
the physical layer of the torus network (see section~\ref{section:tnw} for more
details) and the Ethernet interface.

%- RC
The root card comprises a microcontroller which is operated using
a Linux version optimized for embedded systems. It can be accessed from
a front-end system via its Ethernet interface. The controller and the
service processors of the 16 node cards mounted in the same backplane row
are attached to a shared RS-485 serial bus, thus proving front-end access
to the service processors.  The root card furthermore controls the reset
lines and 2 other serial links going to each of the node cards. The
necessary switching logic has been implemented on a Complex Programmable
Logic Devices (CPLD). A second CPLD is used to implement the second
level of the global signal tree network.

%- SRC
The higher levels of the global signal tree network are implemented on
the superroot cards (see Fig.~\ref{fig:src}) again using a CPLD. Here
a second CPLD is used to implement the logic for the links to the PSUs.
Via these links the PSUs can be monitored and controlled.
To access both devices an Ethernet to serial link converter is used.

%- Backplane
Finally, the last board designed for QPACE is the backplane, a fully passive
board. Its main tasks are power distribution and routing of a large number
of signals (mostly high-speed signals) connecting the node cards among
each other, or node card and root card as well as node cards and cable
connectors. As a result of the compactness of the design a relatively
dense placement of the connectors was required.  In particular, at the
boundary of the board almost all available space is occupied by cable
connectors (see Fig.~\ref{fig:backplane}).

%===============================================================================
\section{Network processor}\label{section:nwp}

A core component of the QPACE architecture is the network processor (NWP).
This device acts as an I/O fabric similar to the chips which on more
standard compute boards connect the processor to external peripheral
devices. While in commodity products these chips are implemented on an
ASIC, we use a Field Programmable Gate Arrays (FPGA), which is a rather
unique feature of the QPACE node card.

% FPGA intro
FPGAs are also semiconductor devices, but they are re-programmable, i.e., the
functional behavior of the device can be changed at any time. This
means that development risks are significantly reduced as the logic can
be modified even after deployment of the hardware. The disadvantage of
using FPGAs is that they are typically slower than custom ASICs and the
price per device is typically higher.%
\footnote{For small projects this disadvantage is typically more than
compensated for by the reduction in development costs.}

An FPGA consists of a large number of programmable \emph{logic blocks}.
A hierarchy of reconfigurable \emph{interconnects} allows to glue these
blocks together in different ways. In some FPGAs so-called hard cores are
embedded which provide a fast implementation of a certain functionality.
For instance, in the QPACE NWP an Ethernet Medium Access Control
(MAC) core is used to implement the lower part of the data link layer
of the Gigabit Ethernet interface.

To define the functional behavior of an FPGA the developer implements a
design using a hardware description language (usually VHDL). Using a complex
tool chain this design is translated into a so-called bitstream which is
used to program the FPGA after power-on of the node card.

% Virtex5 device description
The FPGA which we chose for the QPACE project is the Virtex-5 LX110T
by Xilinx. This FPGA is of medium size (and therefore affordable in price)
with an amount of logic blocks which we estimated to be sufficient for
our design. Furthermore, the number of available general purpose I/O pins
and fast transceivers was large enough to connect the FPGA to the {\cell},
the network PHYs and some other slower devices. As performance was
expected to become a challenge we selected the highest available speed grade
for this device.

\begin{figure}[ht]
\begin{center}
\includegraphics[scale=0.33]{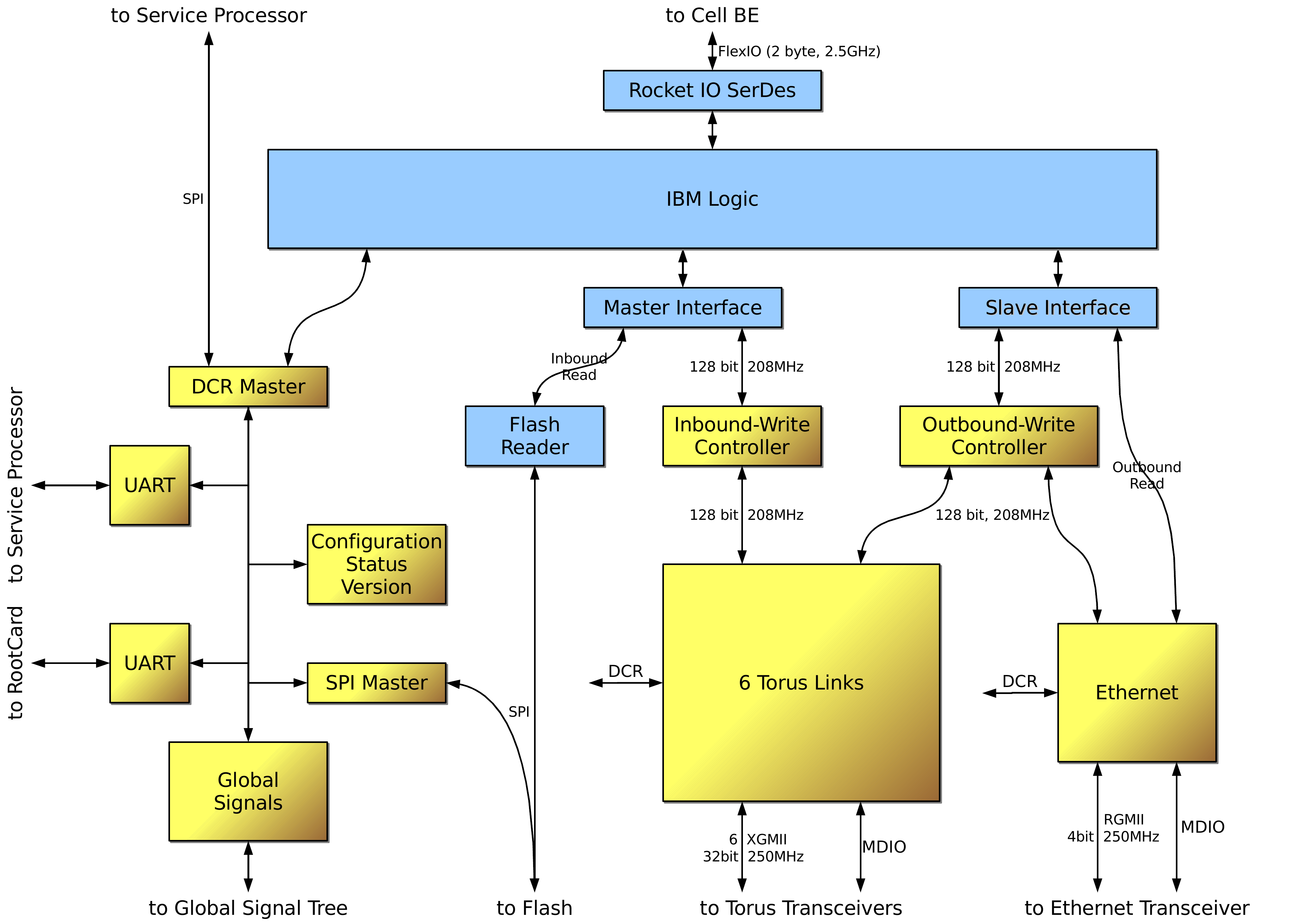}
\end{center}
\caption{\label{fig:nwp}%
Overview of the internal architecture of the Network Processor (NWP).
The blue boxes refer to entities implemented by IBM, the others have
been implemented by the academic partners.}
\end{figure}

% Internal entities overview
In Fig.~\ref{fig:nwp} we give an overview of the internal structure of
our NWP design. The entities depicted in its upper part contain IBM's
implementation of the physical and data link layer towards the {\cell} as 
well as the interface to the application logic designed by the academic
partners. This application logic can be categorized in the following way:
\begin{itemize}
\item Torus network block including 6 links (described in more detail
      in section~\ref{section:tnw}).
\item Gigabit Ethernet interface.
\item Various devices connected to slow interfaces, e.g.,
   \begin{itemize}
   \item 2 serial links connecting the {\cell} to the service processor and
         the root card.
   \item An interface to the global signal tree network.
   \end{itemize}
\end{itemize}

% External lines
With the NWP basically being a large I/O fabric, most entities are connected
to external lines. This includes:
\begin{itemize}
\item Two 8-bit wide (full-duplex) bi-directional high-speed links connecting the NWP to
      the {\cell} with a bandwidth of up to 5~GBytes/sec per direction.
\item Six (full-duplex) bi-directional links to the 10-Gigabit Medium Independent Interfaces\linebreak
      (XGMII) of the torus network PHYs. Per link and clock cycle 32 bits
      can be sent and received at a clock speed of 250~MHz.
\item One common Reduced Gigabit Medium Independent Interface (RGMII) interface
      connecting the NWP to an external Ethernet 1000BASE-T physical
      transceiver. Running at 250 MHz per direction 4 bits per cycle can
      be communicated.
\item A 4-bit interface (2 differential lines per direction) interface to
      the global signal tree network.
\item Lines connected to the 2 Universal Asynchronous Receiver Transmitters
      (UART) which are interfaces to the serial links.
\end{itemize}

% FlexIO link
Major challenges had to be solved to connect the {\cell} and the NWP via
a high-speed link. The link consists of 2 links with
8 lines per direction each.  The transceiver technology used in the processor is
called FlexIO, a technology from Rambus which allows for a speed-up
of data rates up to 8 GHz.  On the other side these lines are connected
to Xilinx RocketIO transceivers which are able to sustain data rates up to
3.75~Gbit/sec. With both technologies not being fully compatible
careful tuning of the design at the physical link layer was required \cite{flexio}. We
have verified that the link can be operated in a stable manner at 2.5~GHz without bit
errors during several hours on hundreds of node cards.  At this speed
the link has a bandwidth of 5~GBytes/sec per direction, which roughly
balances the bandwidth of all 6 torus network links.  Single node cards
have been successfully tested at 3~GHz.

% DCR
While the high-speed network interfaces are connected via a fast data path
to the processor-NWP link, the slow interfaces are accessed via a simple
shared Device Control Register (DCR) bus. This bus is also used to access
the configuration and status registers of the network interfaces.

% Resource usage
In Tab.~\ref{tab:fpgaUsage} we show for some of the most relevant
logic blocks, like Flip Flops, Lookup Tables (LUT) and pins,
as well as some special features, like block memory modules (BRAM), 
how many of them are actually used to fit our design into the given FPGA.
As can be seen from this table the filling fraction of the FPGA is very high.
As a consequence mapping the design taking all relevant timing constraints
into account has become a challenge. We were therefore forced to compromise
on the clock speed. Currently a core part of the design is operated at
166~MHz, which is still significantly below our target speed of 208.3~MHz.

To analyze the FPGA's resource utilization in more detail, we have split
the number of used registers and LUTs according to the design entities.
The results are shown in Tab.~\ref{tab:entityUsage}. About 50\% of these
resources are occupied by the logic which implements the interface to
the {\cell}. The next-to-biggest consumer is the torus network with
about 6\% per link.

\begin{table}[ht]
\begin{center}
\subfloat[Primitives utilization]{
\label{tab:fpgaUsage}
\begin{tabular}{|l|r|r|r|}
\hline
Primitive		& Available	& Used		& Usage \\
\hline
Slices			& 17,280	& 16,029	& 92\% \\
LUT-Flip Flop pairs	& 69,120	& 51,018	& 73\% \\
\hline
Slice registers		& 69,120	& 38,212	& 55\% \\
Slice LUTs		& 69,120	& 36,939	& 53\% \\
BRAM/FIFO		&    148	&     53	& 35\% \\
\hline
Pins			&    680	&    656	& 96\% \\
\hline
\end{tabular}
}\\
\subfloat[Usage per design entity]{
\label{tab:entityUsage}
\begin{tabular}{|l|r|r|r|r|}
\hline
			& \multicolumn{2}{|c|}{Registers}
			& \multicolumn{2}{|c|}{LUTs} \\
\hline
{\cell} interface	& 20,225 & 53\% & 16,915 & 46\% \\
Torus network		& 13,672 & 36\% & 14,252 & 39\% \\
Ethernet		&  1,537 &  4\% &    894 &  2\% \\
Other			&  2,778 &  7\% &  4,878 & 13\% \\
\hline
\end{tabular}
}
\end{center}
\caption{FPGA device utilization:
The upper table compares the utilization of the available logic blocks and
specific features for our current design.
In the lower table we split the number of used blocks and features
according to the different design entities.
}
\end{table}

%===============================================================================
\section{Torus network}\label{section:tnw}

The torus network has been designed to meet the following requirements:
\begin{itemize}
\item Provide connectivity with the six nearest neighbors within a
      torus topology.
\item The network should allow to send and receive messages directly
      from the SPE of one node to an SPE of a neighboring node without
      support from the PPE and without copying the data to main memory.
\item The bandwidth performance goal was set to $O(1)$~GByte/s per link
      and direction. From our performance estimates \cite{lat07,cise} we
      expected this to be sufficiently large such that the network bandwidth
      would not become a bottleneck for our applications.
\item We aimed for a small latency for LS to LS copy operations
      of the order of $O(1)$~$\mu$sec. This latency is small compared
      to typical commodity solutions but larger than for previous custom
      processors where the networking functionality had been integrated on
      the chip (so-called system-on-a-chip designs).
\end{itemize}

To meet these requirements and to ensure that the logic of all 6 links
would fit in our FPGA required an optimized design.

% Communication model
In order to minimize protocol overhead and to avoid handshake between
transmitter and receiver we adopted a two-sided communication model.
When node $A$ initiates transmission of a message to node $B$
it has to rely on node $B$ initiating a corresponding receive operation.
It is the responsibility of the programmer to ensure that the communication
parameters (e.g., message size) and the order of the communication instructions
match.

% Datagrams
As a further simplification we restrict message sizes to be a multiple of
128~Bytes as well as source and destination addresses to be 128~Byte aligned.
This is a natural choice for the {\cell} as for the internal bus, the EIB,
a similar simplification had been made. On the torus network link the messages
are split in packets which consist of a 4~Byte header, a 128~Byte payload
and a 4~Byte checksum. This means that the protocol overhead added at the
data link layer increases the number of transmitted bytes by only 6\%.

\begin{figure}[ht]
\hspace*{5mm}
\subfloat[Data and control paths]{
  \label{fig:tnw}
  \includegraphics[scale=0.35]{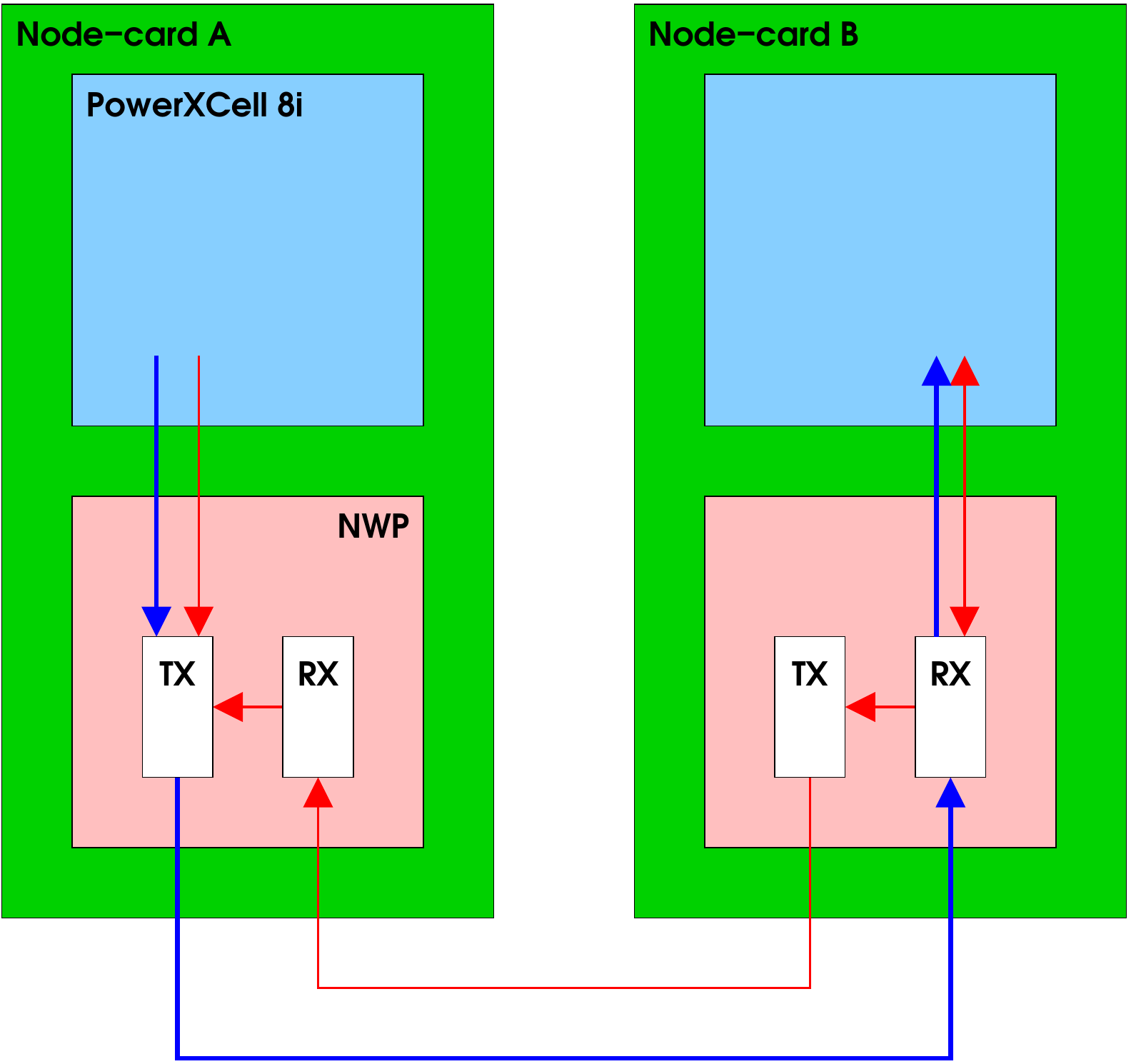}
}
\hfill
\subfloat[Eye diagram]{
  \label{fig:eyes}
  \raisebox{10mm}{\includegraphics[scale=0.35]{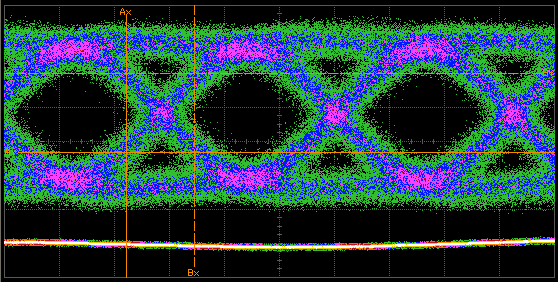}}
}
\hspace*{5mm}
\caption{The left panel shows the data (blue) and control (red) paths for the
torus network.
The right figure shows an eye diagram measured close to a torus
network receiver. For this measurement a link running at 3.125~GHz
was selected which is routed through about 40~cm of PCB, 50~cm cable,
2 PCB-PCB connectors and 2 PCB-cable connectors.
}
\end{figure}

% Transactions and handshakes
The communication of messages between different processors is split into
multiple transactions. In our design for all transactions data is pushed by
the source to the destination.  For each transaction we have to 
ensure that no loss of data occurs. We do this by employing different
mechanisms:
\begin{enumerate}
\item If the source device for a particular transaction initiates a push
      operation the write request may not be acknowledged once no more
      buffer space is available in the destination device, i.e.~any
      further write operation is blocked.
\item The source device sends the data but keeps a copy of each packet
      until receipt of the packet has been acknowledged by the receiver.
      If receipt is not acknowledged, it automatically retransmits the packet.
\item Finally, the source device may be requested to wait until it is
      provided with a credit by the destination device. This credit may
      contain additional information where to write the data.
\end{enumerate}

If any of these transactions is stalled then back-pressure will be generated

% Data integrity
Also data integrity must be preserved during all transactions. In our
design we make the (usual) assumption that the data buses within a
chip are safe, while external links need to be protected by checksums.

% Data/control paths
Within our two-sided communication model the sender starts a send operation
simply by moving a message using a DMA put operation from the attached
storage (e.g., the LS) to one of the transmit (TX) FIFOs%
\footnote{FIFO stands for First In, First Out and describes a storage
entity with a write port and a read port. Data written to this FIFO can
only be read in the same order as they were written.}
in the NWP (see Fig.~\ref{fig:tnw}). There is one such 2~kByte large FIFO
for each of the 6 links. Once the data arrive in this FIFO
the data are queued for transmission and will automatically be sent out
without further intervention of the processor. The
necessary control information is encoded in the address used in the DMA
put operation. When the FIFO is full further writing is disabled, which
will eventually create back-pressure between NWP and SPE.

Once the data is queued for sending, it will immediately be transmitted over
the network link.
Since the receiver may either not be able to accept the
data in its receive buffers or identify data corruption the sender
must keep a copy of the packet until the receiver has returned a 4~Byte
command packet. This packet contains either an \texttt{ACK} (acknowledge)
or \texttt{NACK} (not acknowledge). In the latter case the sender will
automatically retransmit the packet.

So far, all transactions are initiated by the sender. To initiate the last
step, in which data are moved from the NWP of the receiving node to the
processor (or eventually main memory), the sender has to provide a
credit to the receiving link in the NWP. Once data have arrived in the
receive buffer and a matching credit is found the link requests access
to the interface towards the {\cell}. After an arbiter provided a grant,
data are moved to their destination and the processor gets notified once the
operation has completed.

% Virtual channels
As a result of the requirement that any of the SPEs can be an endpoint of
a communication we have to take into account that multiple pairs of
SPEs on neighboring nodes may share the same physical link. We employed the
concept of virtual channels. By using different channels up to 8 pairs
of sender and receiver are logically separated. In the receiver this
requires the instantiation of a second level of arbitration
and logic to reorder data packets from different channels
according to the available credits.

% Addressing
Although protocol overhead has been minimized, there is quite some flexibility
to control the destination address of a network packet. The final destination
address is determined as a sum of 3 addresses:
\begin{itemize}
\item A base address defined by the receiver. It typically points to a
      page in main memory or the base address of the LS and is not
      changed after initialization of the network. One such address
      can be defined per link and virtual channel.
\item A remote offset which is provided by the sender and transmitted
      in the packet header.
\item A local offset defined by the receiver when providing a credit.
      This local offset, e.g., defines the address of the receive
      buffer within the LS relative to the base address.
\end{itemize}

% Physical layer
For the physical layer of the link we decided to use 10-Gigabit Ethernet
(10-GbE). This layer is implemented by external network devices, called
PHYs. Data is moved out of the Network Processor (NWP) in a 32-bit wide
data bus at a speed of 250~MHz. The PHY serializes and encodes the data.
Per link and direction there are 4 serial lines with a data rate of
2.5~GBit/sec each.\footnote{We operate the links at 2.5~GHz instead
of 3.125~GHz used in standard 10~GbE.} Taking the overhead of 10/8-bit
encoding into account, this results in a gross bandwidth of 1~GByte/sec
per link and direction.

Because of the compactness of the system design we could limit the maximum
cable length to 50~cm. This makes the use of simpler and therefore
cheaper cables possible. Based on signal integrity simulations we chose
compact flat cables from Samtec with a width of up to 10~cm and a capacity
of up to 80 pairs of differential signals, i.e., up to 10 links can be
routed via one single cable. In Fig.~\ref{fig:eyes} we show the results
of a so-called eye diagram measurement. Such diagrams show the overlay
of a large number of snapshots sampled with respect to a reference clock.
Only if the signal is stable within a certain time window and if
transitions between logical '0' and '1' occur outside this window
the data will be sampled correctly. A transition inside the window (i.e.,
inside the ``eye'') is likely to result in a bit error.

\begin{figure}[ht]
\begin{center}
\includegraphics[scale=0.70]{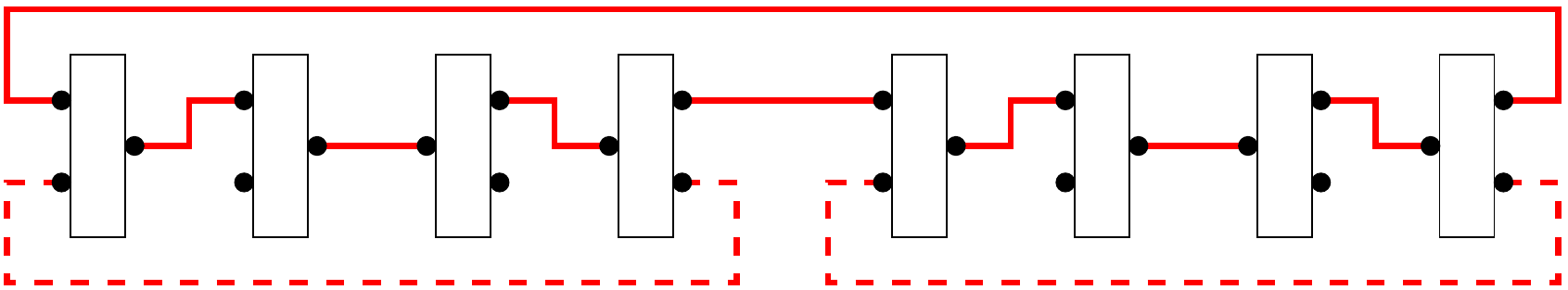}
\end{center}
\caption{\label{fig:redundant}%
The picture shows 8~node cards which by switching between PHY
interfaces in 1~dimension (here: $z$-direction)
can be connected either as 1 ring of 8 or 2 rings of 4 node cards.}
\end{figure}

% Links and machine partitioning
The QPACE node cards are interconnected as a 3-dimensional torus.
The largest partition which fits on a single backplane is of size
$(x,y,z) = (1,4,8)$. To increase the number of nodes in the $y$-direction
cables connecting backplanes in the vertical direction have to be used.
Similarly, in the $x$-direction backplanes are connected in the horizontal direction.
In order to have some flexibility in re-partitioning the machine we make
use of a feature of the network PHYs used on the node cards, which come with
two serial interfaces. Controlled by software it is possible to select either the
primary or the redundant interface. By connecting some of these redundant
interfaces to additional backplane lines it is possible to partition the
machine in different ways. In Fig.~\ref{fig:redundant} we show how this
can be done for a set of 8~nodes connected in the $z$-direction.
Given this flexibility we can partition the machine such that we have
\begin{enumerate}
\item 1, 2, 4, 8 nodes in the $z$-direction.
\item 1, 2, 4, 8 or 16 nodes in the $y$-direction.
\item 1, 2 or $2 N$ nodes in the $x$-direction, where $N = 1, 2, \ldots$
      is the number of interconnected racks. A change of $N$ requires
      re-cabling.
\end{enumerate}

% Performance results and link errors
To measure the performance of our network we use the following benchmarks:
\begin{itemize}
\item A ping-pong type of test was implemented to estimate the latency.
      Here node $A$ sends a message of size 128~Bytes (i.e., 1 packet)
      from its LS to node $B$. Once the data have arrived in the destination LS
      of node $B$ this node returns the packet back to $A$.
      We estimate the latency by measuring the round-trip time on node $A$
      and divide this number by 2.
\item An exchange type of test was implemented to measure the bandwidth.
      In this test node $A$ sends a message to $B$ and vice versa.
      We allow for multiple messages to be sent concurrently.
\end{itemize}

With this setup we found the latency to be about $3~\mu$secs. We also measured
the time needed starting from the point where the packet arrives at the
transmit FIFO until the DMA engine of the receiver link moves the packet
to the processor. Here
we measured a latency of about $0.5~\mu$secs. A large fraction of this
latency is due to the logic in the PHYs (where data encoding and decoding has to
be performed). But the overall latency is largely dominated by the time needed
to move data from the processor to the NWP and vice versa.

\begin{figure}[ht]
\begin{center}
\includegraphics[scale=0.40]{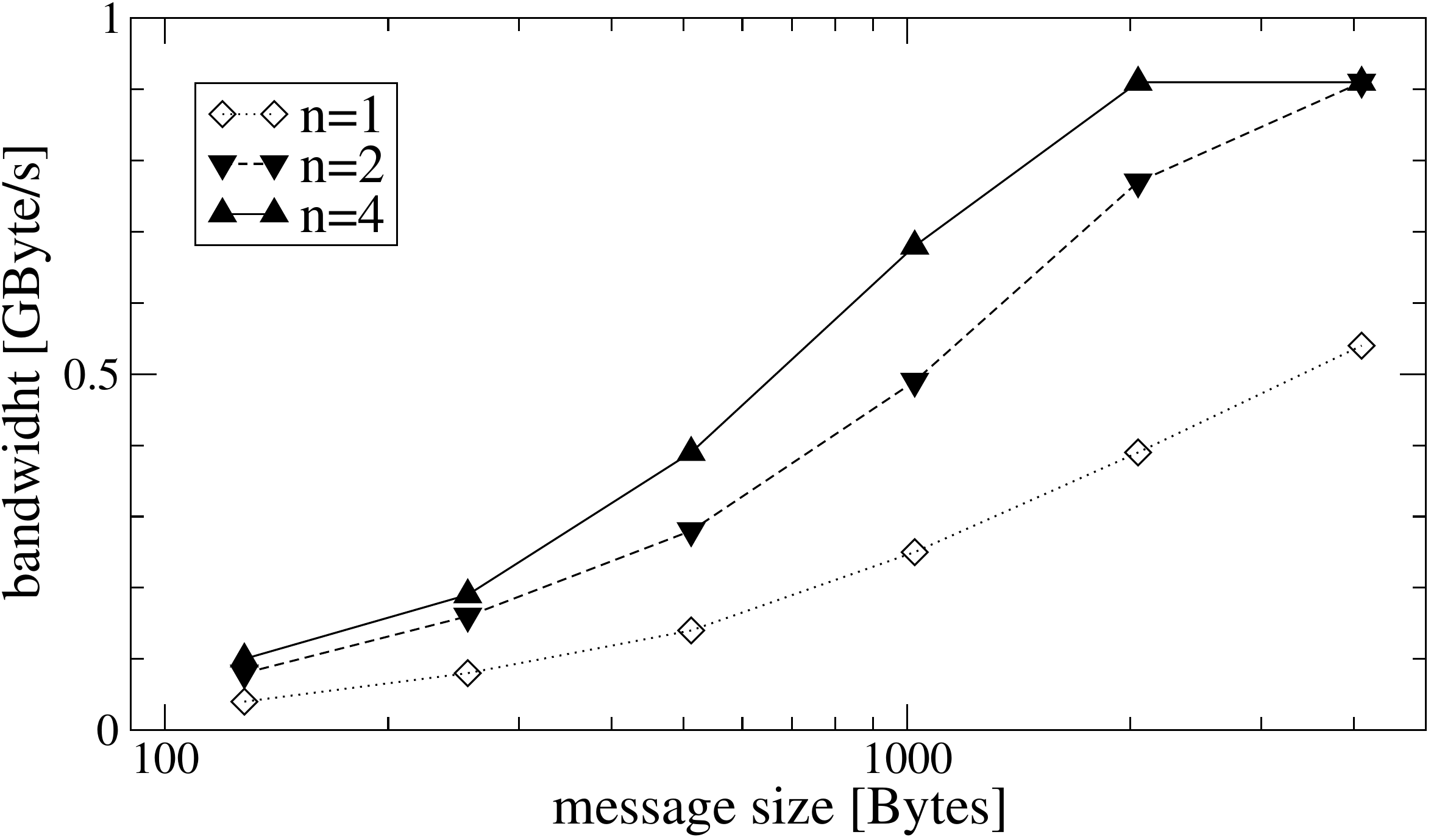}
\end{center}
\caption{\label{fig:tnwBench}%
Bandwidth measured on a single torus
network link as a function of the message size, where $n$ is the number of
messages which are in flight concurrently.}
\end{figure}

The bandwidth depends on the message size and the number of concurrent
communications. We show the results from different measurements in
Fig.~\ref{fig:tnwBench}. As can be seen from this picture it
is possible to get close to the theoretical maximum bandwidth of 0.9
GBytes/sec (taking protocol overhead into account).

%===============================================================================
\section{Cooling and power efficiency}\label{section:system}

One of the biggest challenges for the integration of the node cards into a single
system was the cooling. For all currently operated supercomputers air
is used to remove the heat from heat-sinks (or heat-pipes) connected
to the electronic components which generate the heat, e.g., the
processor. Sometimes liquid cooling is used for heat exchangers inside
the rack to cool the air before it either re-enters the computer or 
leaves the rack. Although this setup is widespread it has its limitations.
Air is per se not very suitable for heat transport. As a result large
amounts of air have to be pushed through the system, which requires a large
amount of space and fans. Increasing the space between the node cards has
impact on the costs, because a larger number of backplanes and racks
is needed as well as longer and therefore higher quality cables have to
be selected.  The required fans have an impact on operation by increasing
operational and maintenance costs due to increased power consumption
and the risk of failure.

On the other hand, liquid cooling is often considered to be risky because
it brings water close to expensive components that will be damaged when
exposed to water. Liquid cooling solutions tend to be expensive and often
have significant impact on maintenance as electronic components may become
difficult to access.

\begin{figure}[ht]
\subfloat[Node cards with coldplate]{
  \label{fig:cooling}
  \includegraphics[scale=0.20]{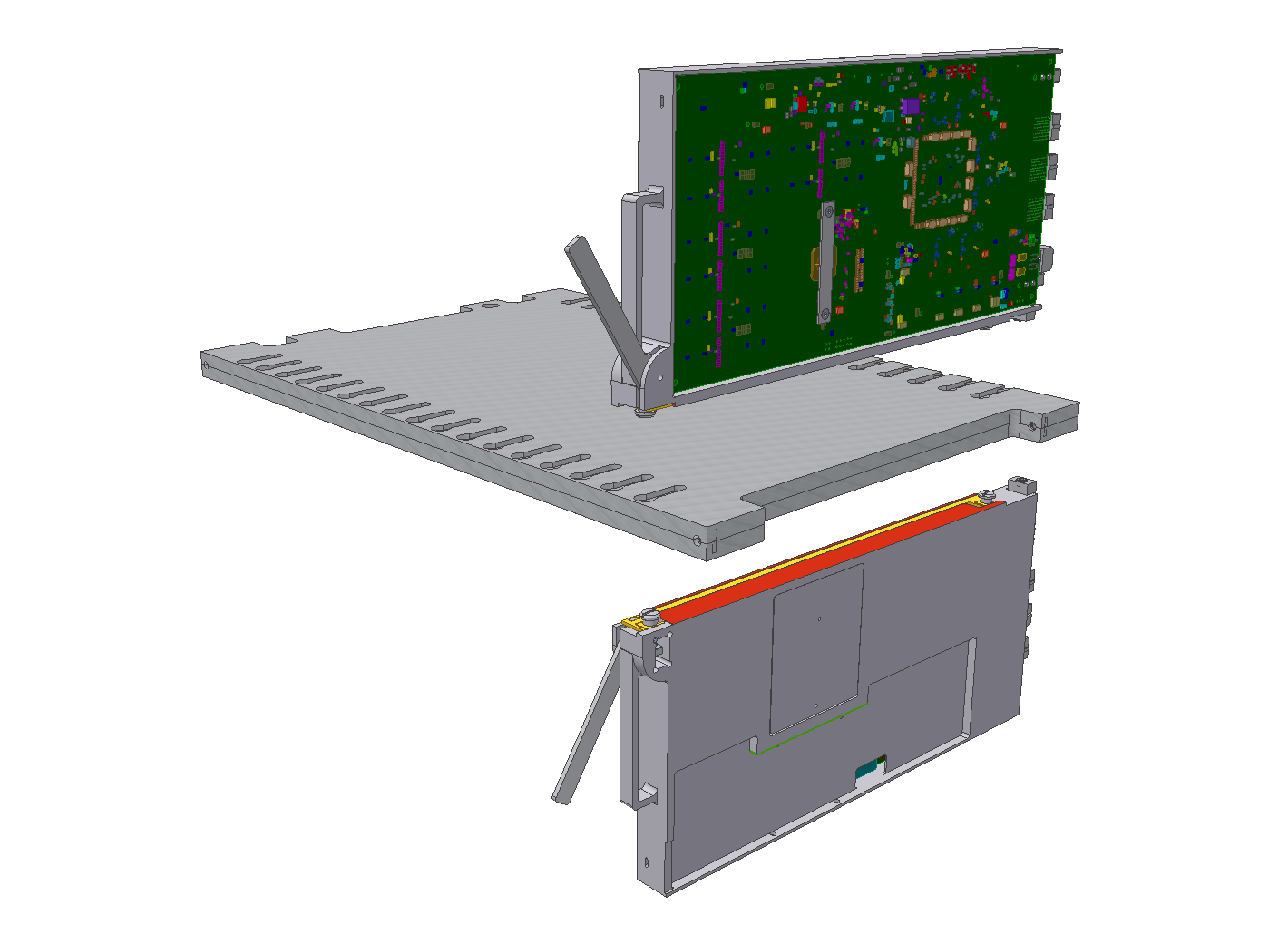}
}
\hfill
\subfloat[Temperature distribution in cold plate]{
  \label{fig:coldplate}
  \includegraphics[scale=0.15]{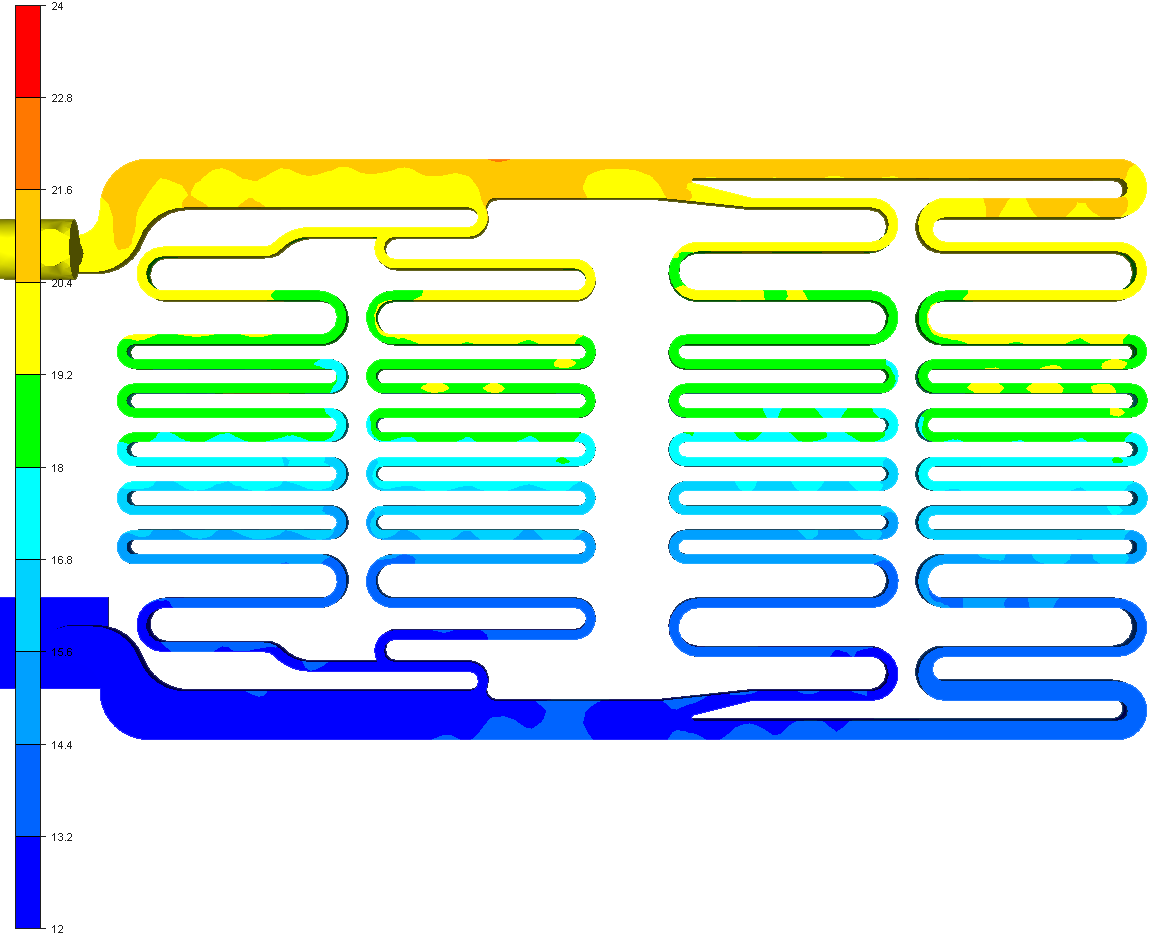}
}
\hspace*{5mm}
\caption{The left panel shows a cold plate together with node cards, one about
to be attached from above, the other from below. The red strip indicates the
thermal interface between node card and cold plate.
Once the node cards are connected to the cold plate the handle attached
to the node card has to be moved into closed position. Springs then press
the node card onto the cold plate.
In the right panel the result of a simulation of the temperature distribution
inside the water-conducting channels of the cold plate are shown.}
\end{figure}

% Cooling concept
The challenge for QPACE was to realize a completely new liquid cooling concept
which would address the typical disadvantages of liquid cooling.
The node cards are packed into thermal boxes made of aluminium. These boxes act
as a single big heat sink conducting the heat from the internal components
to the surface. The thermal box is then connected to a cold plate.
Water is pumped through channels inside the cold plate which moves the
heat out of the system. See Fig.~\ref{fig:cooling} for details.

There is no water flowing through the thermal box and a so-called
dry connection is used to connect the node card with the cooling
system. Therefore, after installation of the machine the water circuit
does not have to be opened for almost any of the expected maintenance
operations.  This eliminates the need for expensive, self-healing valves
often used in other systems.

There are two critical thermal interfaces which have to be kept under control.
The first interface is between the electronic components on the node card
and the thermal box. From inside, the thermal box had to be carefully
designed according to the height of the components and the amount of
heat they generate. Thermal grease is used to establish a good thermal
contact between these components and the aluminum. Particular demanding is
the design near the processor which generates $>50\%$ of the heat on the
board. The other components which need cooling are the memories, the
FPGA, the network PHYs, and the power converters.

The second important thermal interface is between the thermal box and the
cold plate. The heat has to be conducted through a rather small surface
of about 40~cm$^2$. In order to avoid thermal grease, which would be
rather inconvenient during maintenance, a suitable type of silicon oil has
been selected to improve the thermal contact. Using oil, which generates only
a rather thin film, mandated an extremely flat surface on both sides of
the interface. Springs mounted in the thermal box additionally improve
contact by pressing the node card onto the cold plate after the node card
has been mounted.

The cold plate has to be carefully designed in order to make sure that
all 16 node cards mounted from the top side and the 16 node cards mounted
from below are equally well cooled. In Fig.~\ref{fig:coldplate} we show
the results for the temperature distribution within the cold plate from
a simulation. In this picture the water enters the cold plate at the
lower left corner from where it is distributed over the whole length
of the cold plate via a channel with a relatively large diameter. When
the water meanders through the small channels it passes from one end of
the node cards to the other end, slowly heating up. The warm water is
collected in another broader channel and leaves the cold plate at the
upper-left corner of the cold plate.  It is important to notice that
there is only a temperature gradient from bottom to top, from left to
right the temperature is constant, i.e., all node cards are cooled equally
well. The results of this thermal simulation have been confirmed by
temperature measurements using the final system.

% Temperatures
An important feature of the QPACE cooling system is the small temperature
difference $\Delta T$ between the temperature of the water inside the
cold plate and the temperature at the processor. This fact indicates that
the thermal interfaces described above are well under control.
Even at maximum load we measure $\Delta T \lesssim 30-40\;\celcius$.
Taking into account that operation of the processor is specified for
temperatures up to $95\;\celcius$
\footnote{The processor can even sustain higher temperatures without being
damaged, but with no guarantee for correct functional behavior.}
using an inlet water temperature $>30\;\celcius$ is a feasible option.
During test runs we have successfully operated completely populated
backplanes with 32~node cards with a water inlet temperature of up to
$35\,\celcius$ and all node cards running a synthetic benchmark
(called PowerLoad SPU) which maximizes the node cards' power consumption.

At high load the temperature at the outlet is about $5\;\celcius$ warmer
than at the inlet. With an inlet water temperature $>40\;\celcius$ the system
is able to generate hot water in a temperature range which is at the lower limit
to be used for heating and to make free air cooling possible year-round.  Using thermal grease for the thermal
interface between thermal box and cold plate it is in principle possible
to reach higher temperatures.

% Water circuit
The cold plates are part of a closed water circuit with a cooling
station CoolTrans from Kn\"urr. This station regulates the temperature
of the cooling water taking also humidity into account. The latter is
necessary to ensure that the temperature in the water circuit stays
above the dew point.

% Power efficiency, VMIN reduction
While the cooling system is powerful and was designed with sufficient
margins, reduction of power consumption continued to be a major target
of the QPACE project.  The {\cell} processor is already very efficient
in terms of power consumption per floating-point processing power. This
statement is supported by the dominance of the {\cell}-based systems in
the top positions of the Green500 list~\cite{green500-11}.

For the QPACE node cards a further reduction of power consumption could
be achieved by tuning the processor voltage for each node card individually.
Using the already mentioned synthetic benchmark PowerLoad SPU we measured
on a set of 32 node cards from the production batch a reduction by
12\%. For this benchmark the average power consumption is about 115~W
per node card. Real applications will not reach the same
density of operations as a synthetic benchmark. We therefore expect the
average power consumption per node card to be $O(100)$~W.

To demonstrate the power efficiency of the QPACE architecture, the HPL
benchmark has been ported to QPACE~\cite{hplqpace}. This project took
advantage of the fact that the Network Processor is reconfigurable. The
machine can thus still be optimized for other applications. In this
particular case an additional DMA engine has been implemented to improve
main memory to main memory communication controlled by the PPE. Based
on this setup it has been possible to achieve a performance of 723
MFlops/W. This put QPACE in the number one spot on the Green500 list,
with a 35 \% improvement compared to the previous number
one~\cite{green500-06}.

%===============================================================================
\section{Operating system, I/O network and front-end integration}
\label{section:frontend}

% Boot process, Linux
Once the service processor has powered on the {\cell}, the processor
starts executing the firmware, which provides a glue layer between hardware
and the operating system. Like for other {\cell}-based systems
the Slimline Open Firmware (SLOF) is used~\cite{slof}. The firmware
boots the node card into Linux.
We are using a standard Fedora distribution with a more recent, slightly
modified kernel. The QPACE proprietary devices required the addition
of a few additional kernel modules (some of them have become part of the
standard Linux kernel). This includes drivers for Ethernet, UART, torus
network and global signals.

\begin{figure}[ht]
\begin{center}
\includegraphics[scale=0.4]{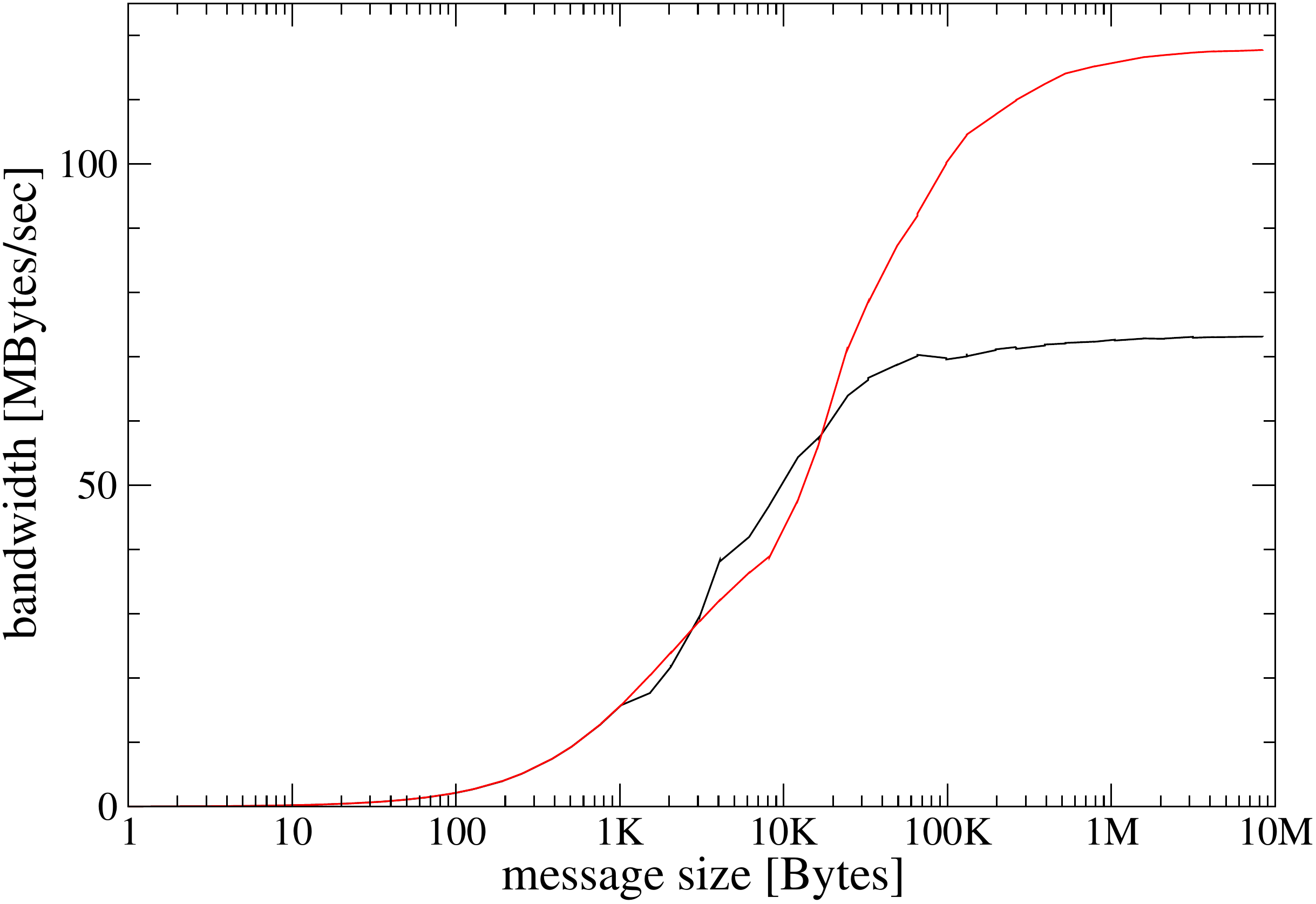}
\end{center}
\vspace*{-5mm}
\caption{\label{fig:eth2}%
Ethernet bandwidth in units of MBytes/sec measured on a link connecting
2 node cards as a function of the message length. The red curve has
been obtained using jumbo frames.}
\end{figure}

% Ethernet network
The Linux operating system is loaded via the Gigabit Ethernet network.
With the data link layer as implemented on the QPACE node card it
is possible to saturate the bandwidth provided by the underlying
physical link layer.
Evaluations performed within a QPACE rack have shown a node-to-node
ping round-trip time of $50$~$\mu$sec and a sustained TCP bandwidth of
0.1237~GBytes/sec in jumbo frame mode. In this mode the theoretical maximum
bandwidth of Gigabit Ethernet is 0.1239~GBytes/sec.
Results from these benchmark measurements are shown in Fig.~\ref{fig:eth2}.
Note that this bandwidth will not be achieved once a larger number of
nodes is involved due to limited capabilities of the Ethernet switches.

% Switch tree
All node cards within one rack are connected to 6 first-level switches.
Per rack there are 24 Gigabit Ethernet uplinks connecting the first-level
to the second-level switches, which are part of the front-end system.
An external bandwidth of $O(2-3)$~GBytes/sec per rack is sufficient for
non-I/O intensive lattice QCD applications, like the generation of
gauge fields with dynamical fermions. For more bandwidth it would be
necessary to upgrade the switches (e.g., by adding 10-GbE uplinks).

% Front-end description
The front-end systems of the currently deployed QPACE installations
consist of 8 IBM System x Linux servers. A login server acts as
central user access point where users can login to submit jobs and manage
their data. The so-called master server acts as central system control and
monitoring instance. All other servers will be used to provide fast disk
storage access. The number of disks attached to these servers is sufficient
to sustain a bandwidth of up to 2 GBytes/sec.

% Lustre
We plan to deploy a Lustre parallel file system as soon as (in the near
future) clients will become available for the Linux kernel currently
used for QPACE.

%===============================================================================
\section{Application code and performance}\label{section:appl}

% Why CBE is difficult
The first and main challenge of porting application code to QPACE is
related to the {\cell} architecture, which is quite different from
commonly used x86 and PowerPC processors causing an obvious lack of
portability. The difficult part is the implementation of the
application on the Synergistic Processing Elements (SPE).%
\footnote{It is easy to port applications to the Power Processing Element (PPE),
but high performance can only be reached when using the SPEs.}
The programmer herself has to take care of data management,
i.e., allocation of Local Store (LS) space and explicit
DMA get and put operations to load data from or store data to main memory.

% Optimisation
For optimizing code on QPACE one faces the following challenges:
\begin{itemize}
\item With essentially all lattice QCD kernels being memory bandwidth
      limited it is mandatory to reduce the number of memory accesses
      to an absolute minimum and to optimize re-use of data in the on-chip
      LS.
\item When parallelizing the application communication has to be organized
      in such a way that the network latencies of $O(10,000)$ clock cycles
      are hidden.
\item Efficient use of the SPU's floating-point pipeline, i.e.,
      SIMDization of the code.
\end{itemize}

% Block, DD
When implementing a solver, where performance is completely dominated by
the performance of the fermion matrix times vector multiplications, an
obvious strategy is to choose a data layout which allows for a processing
of single blocks. One such strategy is described in \cite{lat07}. A more
natural approach %, which will be described in more detail in~\cite{nobile},
is based on the Domain Decomposition algorithm~\cite{Luscher:2003vf}. This
algorithm has the advantage that a very high performance can be achieved
at reasonably large local lattice size. A sustained performance of
$O(45)\%$ has been measured for the internal block solver. Mainly due
to the required memory access the overall solver performance drops to
$O(20)\%$ (with room for further improvement).

% Other attempts
Some groups which reported on porting lattice QCD kernels or full lattice QCD
applications to the Cell processor either avoided some optimization problems,
e.g., by restricting themselves to the on-chip memory \cite{Spray:2008nt},
or have limited their efforts to optimize their data layout
\cite{Shi:2009uq}. An analysis of different optimization strategies can
be found in \cite{rennes}.

% Projects overview
The main purpose of QPACE is to provide the compute power needed to generate
gauge configurations with dynamical fermions. We are currently working on
application programs using Clover fermions and domain wall fermions.
But there are also attempts to port, e.g., multi-grid methods \cite{multigrid}
to QPACE. Furthermore, a 2D code for multi-phase
fluid dynamics based on the Lattice Boltzmann approach has been written
and validated on a 32 node QPACE system, and a (DP) efficiency of $\simeq
20\%$ of peak has been reached. Further performance optimizations are
possible. A preliminary report of this activity is in\cite{boltzmann}.

%===============================================================================
\section{Project organization}\label{section:prjorg}

% Organisational models
Within the QPACE project a new model of collaboration that was new to all
partners has been developed. In the QPACE setup we have on the one hand
a set of academic partners and on the other hand one strong industrial
partner, IBM. Some of the academic institutions as well as IBM had
a double role of being partner and customer/vendor at the same time.

% Team
While the customer/vendor model was particularly relevant during the
manufacturing process in order to define clear responsibilities, the
partnership model has important advantages for the development process.
A joint development team was able to quickly adjust and refine the
requirements and to bring together the know-how and skills to react quickly
and flexibly to all the problems which needed to be addressed and solved.

% Time table
The project has been carried out in an extremely short period of time.
Officially the project started in January 2008. At this point a number of
major design decisions had already been made and first experiments
addressing key technological challenges had been performed.
Already in June 2008 the first versions of the node card became available,
prototypes of backplane and root card followed in July 2008. In December
2008 and January 2009 extensive hardware integration tests have
been performed, which where mandatory to decide on release for
manufacturing. All components where finally released for manufacturing in
February and March 2009. In April 2009 the first components started to
arrive from the manufacturers and the machine integration phase started.
After pre-integration of the racks at the University of Regensburg, the
racks were shipped to the IBM B\"{o}blingen lab.
There all racks were first fully assembled and tested before being
deployed at the installation sites. Deployment both at the
J\"{u}lich Supercomputing Centre and the University of Wuppertal has been
completed in August 2009 and a test operation phase has started.

%===============================================================================
\section{Conclusions and outlook}\label{section:conclusion}

QPACE is a new, scalable parallel computer based on the {\cell}
processor which is optimized for lattice QCD applications. While a
rack consists mainly of custom designed components, the compute
power is provided by commodity processors which are interconnected
by a custom network. Key parameters of a QPACE rack are collected in
Table~\ref{tab:rack}.

% Rack parameters
\begin{table}[t]
\begin{center}
\begin{tabular}{|l|l|}
\hline
Peak performance single/double precision & 26/52 TFlops \\
\hline
Rack size (w$\times$d$\times$h) & $80\times 120\times 250$ cm$^3$ \\
\hline
Weight & $O(1,000)$~kg \\
\hline
Maximum power consumption & $< 35$ kWatts \\
\hline
Average power consumption & $O(29)$ kWatts \\
\hline
\end{tabular}
\end{center}
\caption{\label{tab:rack}%
Key parameters of a QPACE rack.}
\end{table}

% Technical highlights
Within this project we tried to explore new technical concepts. The
design highlights of the QPACE architecture include:
\begin{itemize}
\item An I/O fabric has been implemented on an FPGA which is
      directly attached to the {\cell} processor.
\item The processors are interconnected by a fast lattice QCD optimized,
      low latency torus network.
\item A custom system design makes it possible to mount 256 node cards
      within a single rack, thus providing a very high compute density.
\item The compactness of the design is facilitated by a
      novel, cost-efficient liquid cooling system.
\item The power consumption of the node cards has been further reduced
      in the course of this project making QPACE one of the most
      power-efficient currently available parallel computers.
\end{itemize}

% Installations
The QPACE hardware has been deployed at two installation sites,
the J\"{u}lich Supercomputing Centre and the University of Wuppertal. The
installations consist of 4 racks each, i.e., an aggregate peak performance
of 200/400~TFlops (double/single precision).

During burn-in of the hardware
in the integration phase a few defective components had been found.
After deployment at the installation sites and since start of the test
operation phase almost no defects occurred.

\begin{figure}[ht]
\begin{center}
\includegraphics[scale=0.6]{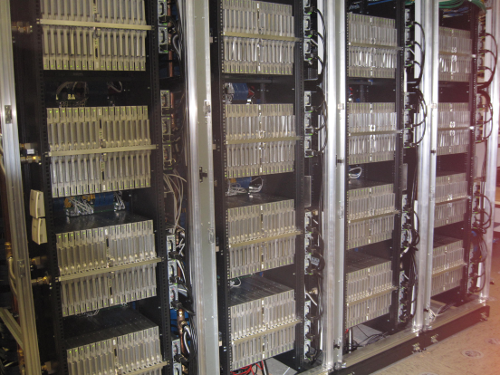}
\end{center}
\caption{\label{fig:jsc}%
4 QPACE racks installed at the J\"ulich Supercomputing Centre.
}
\end{figure}

% Review
While there are still some ongoing development activities, as of
today no serious flaws have been identified which could be attributed
to wrong design decisions. The technical concepts behind the QPACE
architecture have essentially been proven to be successful. One
example is the use of FPGAs for implementing a custom network to
interconnect commodity processors. However, this particular example
also shows that there are non-trivial technical challenges which have
to be solved. With increasing resource usage it becomes more and more
difficult to fit the design into the FPGA. This area still requires
further attention.

% Software and outlook
Currently application codes are being ported to the new architecture.
For a key application kernel, a solver for Clover-fermions based on the
Domain Decomposition algorithm, a sustained performance of
$O(20)\%$ has been reached. As there is room for improving this performance
to up to $O(30)\%$ we hope for a (single-precision) sustained performance
of 10-16~TFlops per rack.

We plan to reach production mode by end of 2009/beginning of 2010. This would
put us in a position to present physics results generated on QPACE at the
next Lattice Symposium.

%===============================================================================
\section*{Acknowledgements}

We acknowledge the funding of the QPACE project provided by the Deutsche
Forschungsgemeinschaft (DFG) in the framework of SFB/TR-55 and by IBM. We
furthermore thank the following companies who contributed significantly
to the project in financial and/or technical terms: Axe Motors (Italy),
Eurotech (Italy), Kn\"urr (Germany), Xilinx (USA) and Zollner (Germany).

%===============================================================================

\end{document}